\documentclass[final,times,twocolumn]{elsarticle}
\usepackage{lineno,hyperref}
\modulolinenumbers[1]
\usepackage{graphicx}  
\usepackage{dcolumn}   
\usepackage{bm}        
\usepackage{verbatim}   
\usepackage{textcomp}
\usepackage{amssymb}
\usepackage{amsmath}
\usepackage{epsfig}
\usepackage{tabularx}
\usepackage{xcolor}

\journal{NIM A}

\begin{document}

\begin{frontmatter}

\title{Simulations and analysis tools for charge-exchange $(d,{}^{2}\text{He})$ reactions  in inverse kinematics with the AT-TPC}

\author[frib]{S.~Giraud}
\ead{giraud@frib.msu.edu}
\author[frib]{J.C.~Zamora }
\ead{zamora@frib.msu.edu}
\author[frib,msu]{R.G.T.~Zegers}
\author[sc]{Y.~Ayyad}
\author[frib,msu]{D.~Bazin}
\author[frib,msu]{W.~Mittig}
\author[frib,msu]{A.~Carls}
\author[frib,msu]{M.~DeNudt}
\author[frib,msu]{Z.~Rahman}
\address[frib]{Facility for Rare Isotope Beams, Michigan State University, East Lansing, Michigan 48824, USA}
\address[msu]{Department of Physics and Astronomy, Michigan State University, East Lansing, Michigan 48824-1321, USA}
\address[sc]{IGFAE, Universidade de Santiago de Compostela, E-15782 Santiago de Compostela, Spain}

\begin{abstract}
Charge-exchange $(d,{}^{2}\text{He})$ reactions in inverse kinematics at intermediate energies are a  very promising method to investigate  the Gamow-Teller transition strength in unstable nuclei.  A simulation and analysis software based on the \textsc{attpcroot} package was developed to study these type of reactions with the active-target time projection chamber (AT-TPC). The simulation routines provide a realistic detector response that can be used to understand and benchmark experimental data. Analysis tools and correction routines can be developed and tested from simulations in \textsc{attpcroot}, because they are processed in the same way as the real data. In particular, we study the feasibility of using coincidences with beam-like particles to unambiguously identify the $(d,{}^{2}\text{He})$ reaction channel, and to develop a  kinematic fitting routine for future applications. More technically, the impact of space-charge effects in the track reconstruction, and a possible correction method are investigated in detail. This analysis and simulation package constitutes an essential part of the software development for the fast-beams program with the AT-TPC.

\end{abstract}

\begin{keyword}
AT-TPC\sep 
charge-exchange reactions \sep
simulation \sep
$(d,{}^{2}\mathrm{He})$
\end{keyword}

\end{frontmatter}


\section{Introduction}

During the last few decades, Time Projection Chambers (TPC) have been successfully used as large-volume tracking detectors in many particle physics experiments, e.g., TOPAZ  \cite{KAMAE1986423}, STAR \cite{ACKERMANN1999681} or ALICE \cite{ALME2010316}. However, in the recent years, the operation of TPCs in Active Target (AT) mode have gained a great interest in the nuclear physics community. These type of  devices allow the use of a target medium as a tracking-detection system with a large solid angle coverage and low-energy detection thresholds, which make them ideal for experiments with rare-isotope beams in inverse kinematics. \par

\begin{figure}[!ht]
\centering
\includegraphics[width=0.5\textwidth]{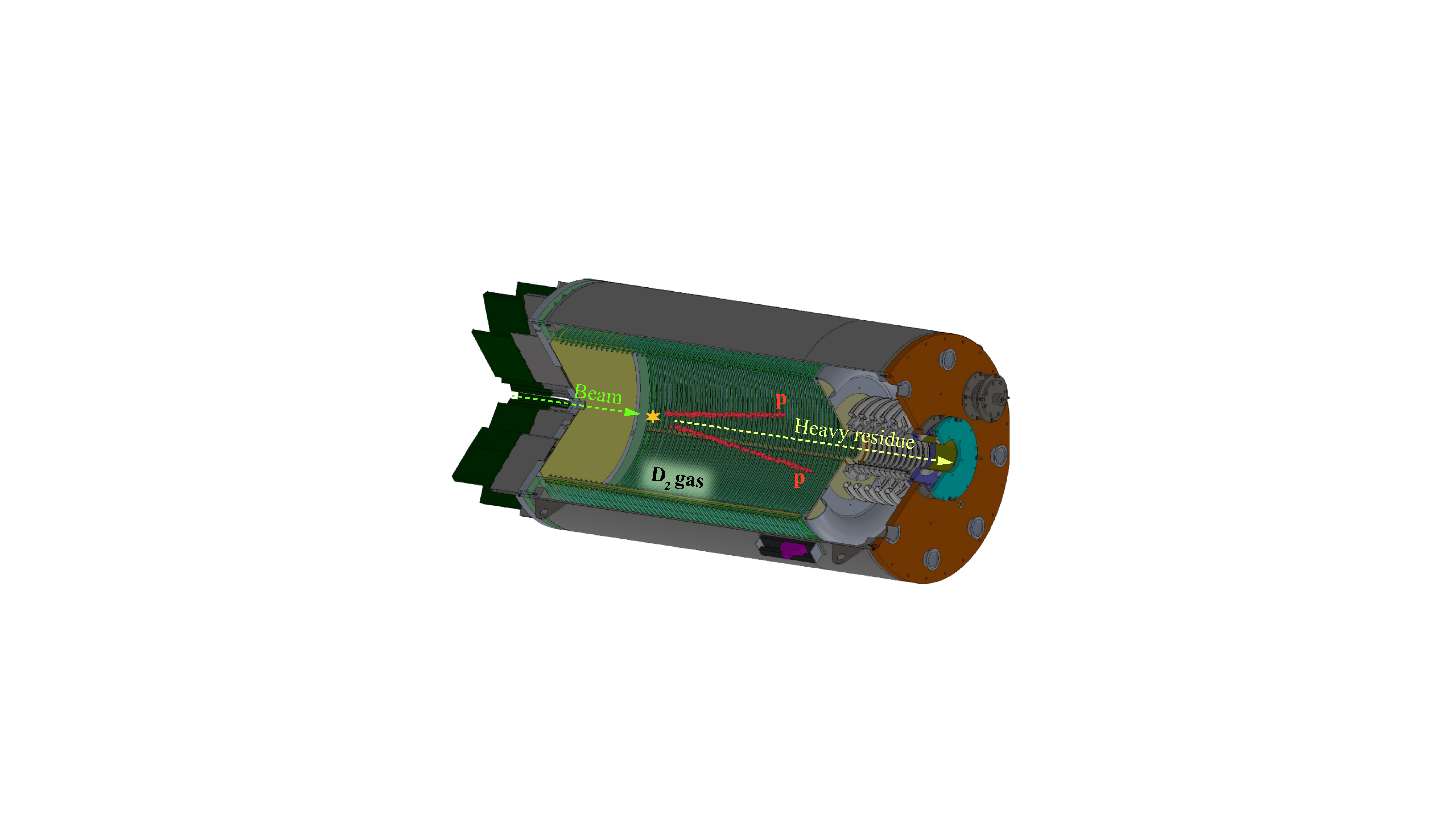}
\caption{\label{attpc}  Schematic view of the AT-TPC with a $(d,{}^{2}\text{He})$ reaction. The active volume, filled with a pure D$_2$ gas, is enclosed by a cylindrical chamber of 100~cm long with a diameter of 50~cm. The beam impinges along the central axis.  The Micromegas plane (upstream) and the cathode plane (downstream) have a hole in the central region that allows the beam to enter and the heavy residue to exit the AT-TPC.  }
\end{figure}

Currently, many facilities around the world are investing a
great effort in the development of active target TPCs as a fundamental part for future research programs 
\cite{HEFFNER201450,FURUNO2018215,MAUSS2019498, SHANE2015513, BRADT201765, KOSHCHIY2020163398}. One of these projects is the already operating Active-Target Time Projection Chamber (AT-TPC)  \cite{BRADT201765} at the Facility for Rare Isotope Beams (FRIB). The AT-TPC has  a cylindrical geometry with dimensions of  100~cm $\times$ \o~50~cm . The gas volume is enclosed by a cathode plate and a Micromegas electron amplifier plate, as shown in  Fig.~\ref{attpc}. The system is designed to take the beam particles impinging along the symmetry axis of the active volume and to detect the  nuclear reactions induced in their path. The beam particles and the reaction products ionize gas atoms while traversing the active volume and generate electrons. Upon applying an uniform electric field, ionization electrons produced by charged particles along their  tracks drift towards  the  Micromegas sensor plane at a constant velocity. The high segmentation of the Micromegas plane ($\sim10^4$  triangular pads) makes possible to obtain the energy loss and a two-dimensional image  of the track, while the third dimension (along the beam axis) is taken from the drift time of the electrons. A hole (\o~3~cm) in the central part of the Micromegas plane creates an insensitive region around the beam axis that allows relatively high beam intensities (see Fig.~\ref{attpc}). This is important for performing experiments in combination with a magnetic spectrometer that enables measurements of heavy residues in coincidence with the tracks in the AT-TPC. The active region of the AT-TPC is separated from the vacuum of the beam line and the spectrometer by 12~$\mu$m polyamide windows.  

Charge-exchange $(d,{}^{2}\text{He})$  reactions at intermediate energies are well known  to be a strong Gamow-Teller (GT) transition filter ($\Delta T, \Delta S=1$) from experiments in forward kinematics \cite{PhysRevC.47.648,OKAMURA19951,PhysRevC.52.R1161,PhysRevC.65.044323,Frekers2004}.  The $(d,{}^{2}\text{He})$ probe in inverse kinematics is a very promising method to study the  GT transition strength  in  far-from-stability nuclei.  The output channel, ${}^2$He, is an unbound system that decays into two protons. In order to achieve this type of experiments with unstable beams, it is necessary to operate the AT-TPC with a pure deuterium gas, which is  used as a target and a tracking medium for the two protons originating in the output channel of the $(d,{}^{2}\text{He})$  reactions.  \par

In this work, we have developed a  simulation and analysis software based on the \textsc{attpcroot} package \cite{Ayyad_2017,ATTPCROOT_git} to study  $(d,{}^{2}\text{He})$ experiments with the AT-TPC.  The simulations offer an ideal method to benchmark and to understand the experimental data taken with this system. The realistic detector response provided by the simulations enables a testing method for  the analysis tools that are applied to the experimental data.  This paper is organized as follows: in Section 2, we give an overview of the simulation routines implemented in this work, Section 3 shows a few remarkable advantages of using coincidences with beam-like particles, space-charge effects and a correction method are discussed in Section 4, and  Section 5  presents the conclusions.

\section{Simulation of $(d,{}^{2}\text{He})$ reactions }

 Simulation routines of $(d,{}^{2}\text{He})$ reactions were developed using the  \textsc{attpcroot} package \cite{Ayyad_2017,ATTPCROOT_git} written in the \textsc{C++} programming language. \textsc{attpcroot} uses the \textsc{fairroot} framework \cite{fairroot} which was developed for the analysis of experiments at FAIR and later redesigned for other  experimental setup. The \textsc{attpcroot} package provides consistent analysis tools  for both simulated  and experimental data. The simulation part of this package comprises three main stages: event generator, digitization and reconstruction. The latter stage is also common for the data analysis part of the package.

\subsection{Event generator}
The simulation events are generated with  the Virtual Monte-Carlo (VMC) package \cite{H_ivn_ov__2008}  that serves as an interface to the \textsc{geant4} toolkit \cite{AGOSTINELLI2003250}. The  sensitive volume of the AT-TPC is defined as a cylindrical geometry filled with a  deuterium   gas (D$_2$) that serves simultaneously as  target and tracking medium for the charged particles involved in the reaction.  Thus, the event generator class creates the particles that are transported by \textsc{geant4} and stores the hit information in a class that handles the simulated data points.\par

\begin{figure}[!ht]
\centering
\includegraphics[width=0.5\textwidth]{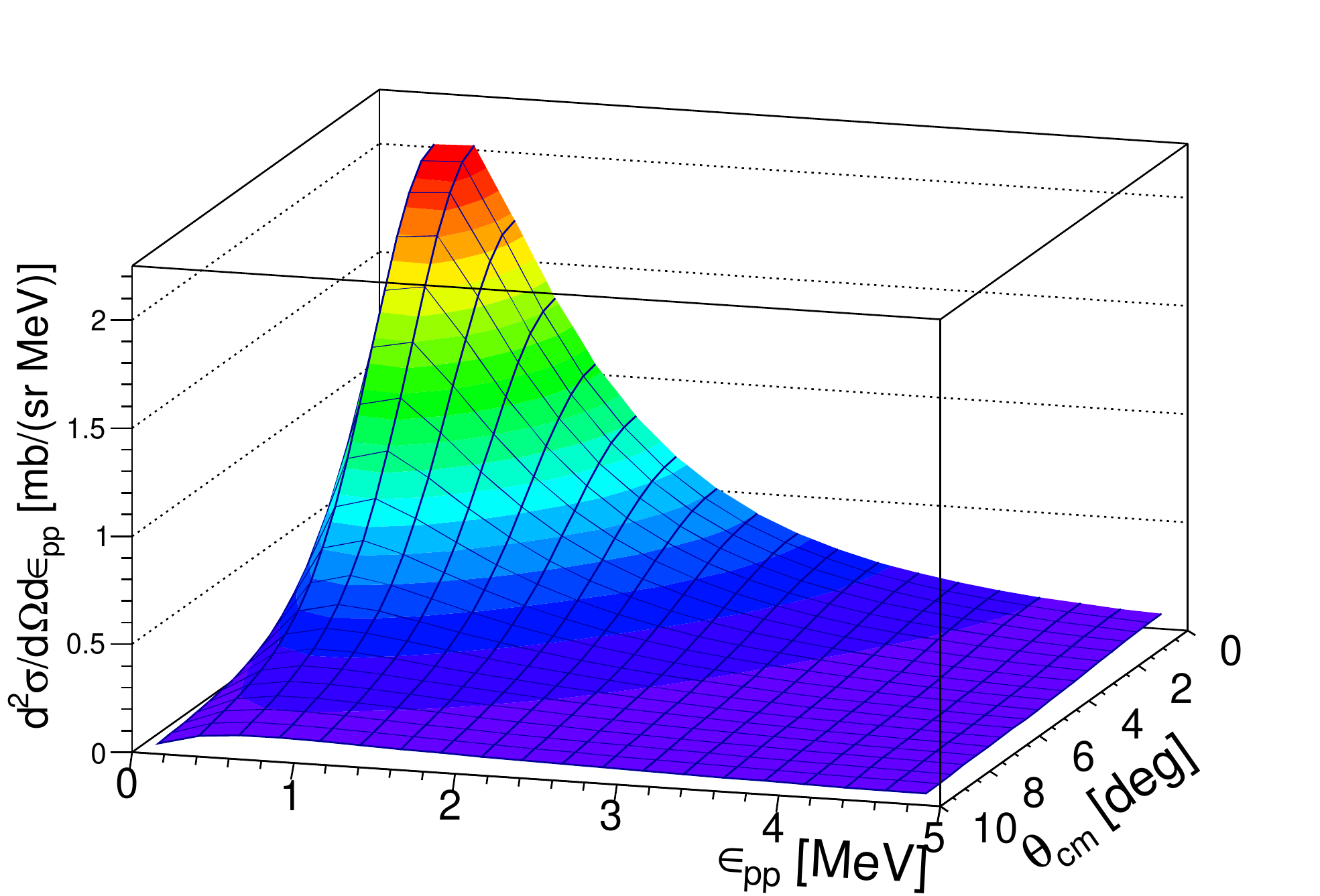}
\caption{\label{dcs} Double-differential cross section of the reaction ${}^{14}\text{O}(d,{}^{2}\text{He}){}^{14}\text{N}$ at 100~MeV/u (populating the ${}^{14}$N $1_1^+$ state at 3.9~MeV) as a function of the scattering angle ($\theta_{\text{c.m.}}$) and the relative energy ($\varepsilon_{pp}$). This cross section was calculated using the code \textsc{accba} \cite{PhysRevC.60.064602}.   }
\end{figure}

The $(d,{}^{2}\text{He})$ events are generated along the active volume using a realistic beam emittance based on the ion-optical properties of the beam line at the S800 spectrometer at FRIB \cite{BAZIN2003629}. Three particles are produced in the final channel of each reaction: two protons  (from the ${}^{2}\text{He}$ decay) and an ejectile. A routine to simulate the  in-flight decay of the residues after the reaction is also included in the code. In order to have  a consistent treatment of the $(d,{}^{2}\text{He})$  reaction, a relativistic three-body kinematics is used to generate the particles in the output channel. An important  part of the $(d,{}^{2}\text{He})$ kinematics is the description of the relative energy between the two protons ($\varepsilon_{pp}$) in the 
${}^{2}\text{He}$ frame. $\varepsilon_{pp}$ is calculated with the code \textsc{accba} (Adiabatic Coupled-Channels Born  Approximation) \cite{PhysRevC.60.064602}, which has been successfully used to reproduce $(d,{}^{2}\text{He})$ data from experiments in forward kinematics \cite{PhysRevC.65.044323,PhysRevC.71.024603,Grewe2004}. Fig.~\ref{dcs} shows the double-differential cross section for the ${}^{14}\text{O}(d,{}^{2}\text{He}){}^{14}\text{N}$ reaction (populating the ${}^{14}$N $1_1^+$ state at 3.9~MeV) as a function of $\varepsilon_{pp}$ and $\theta_{\text{c.m.}}$ (center-of-mass angle) calculated with the code \textsc{accba}. As expected for a GT transition, the differential cross section peaks at forward scattering angles. The relative energy between the two protons  $\varepsilon_{pp}$ is distributed from 0 up to 5~MeV. These angular and energy distributions are included as an input to the event generator to accordingly determine the detection efficiency as a function of $\theta_{\text{c.m.}}$ and $\varepsilon_{pp}$.

\subsection{Digitization}
\begin{figure}[!h]
\centering
\includegraphics[width=0.5\textwidth]{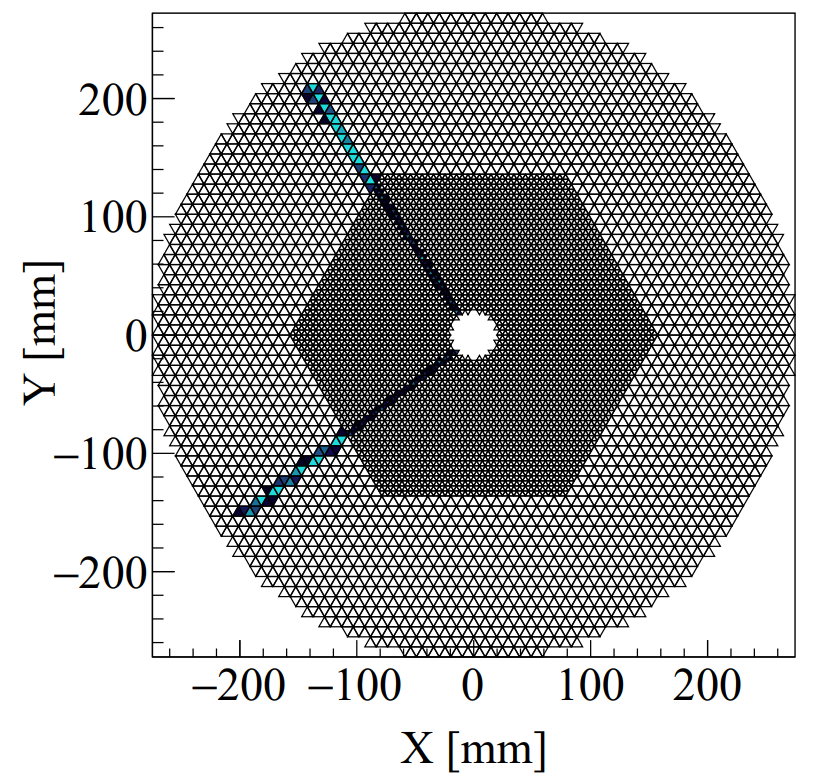}
\caption{\label{tracks} A simulated $(d,{}^{2}\text{He})$ event with the AT-TPC. Note that the pad plane has a 3~cm aperture in the central region that allows the beam-like particles  pass through the active volume.   }
\end{figure}

The digitization stage uses the generated tracks and simulates the detector response.  This part is  divided in two subroutines:  electron diffusion and electronics response. In the electron diffusion subroutine, the number of ionization  electrons is calculated  for every hit based on the energy deposition and the average ionization energy for the gas target. Subsequently, the drift time of the electrons to reach the pad plane is calculated using a drift velocity obtained from \textsc{magboltz} \cite{BIAGI1989716}, which simulates the electron transport in a certain gas  under the influence of an electromagnetic field. It is important to mention that the uncertainty on the reconstructed excitation energy of the ejectile depends on the electron diffusion in the gas volume. For example, a large diffusion parameter  produces thicker proton tracks which degrades the fitting precision.  The corresponding parameters were set to realistic values achieved with the AT-TPC operated  with a pure D$_2$ gas. Effects on the electron drift due to field distortions can be simulated in detail by using electron transport routines in this method, as explained in Section~\ref{spacechar}. Fig.~\ref{tracks} shows an example of a simulated $(d,{}^{2}\text{He})$ event and its respective projection on the pad plane. Each proton hit produces an electron cloud that drifts in the gas volume towards the  detector pixels in the pad plane. In particular, measurements of proton tracks in the AT-TPC require a high gain in the electronics, which can enhance also the detection of $\delta$-rays. The output of this subroutine is coupled to the electronics response module that simulates a signal for each pad. In this part, an electron avalanche generates a pulse shape  by using a realistic response of the Micromegas and the GET (general electronics for TPCs) \cite{POLLACCO201881} system.

\subsection{Reconstruction \label{reco} }
The event reconstruction and further tasks are the same for both simulated and experimental data. After the digitization, a point cloud (collection of hits in the 3D space) for each reaction in the gas volume is extracted. The analysis of these point clouds is performed with  a pattern recognition and fitting routine that searches  $(d,{}^{2}\text{He})$  events and extracts the kinematical information such as  scattering angle, momentum, and excitation energy. 
Tracking algorithms allow to identify the points that belong to the particle tracks (inliers) within a entire point cloud  and reject the noise points (outliers). A tracking algorithm based on \textsc{ransac} (RANdom SAmple Consensus) has been successfully employed to analyze data from the AT-TPC \cite{AYYAD2018166}. However, \textsc{ransac} is rather sensitive to the inlier-outlier threshold and it requires a prior parameter fine tuning  each time for a specific application. Also, a common problem of the algorithm is that it fails when describing data that contain multiple structures, such as point clouds with several tracks that have a significant amount of pseudo-outliers. A few improved versions of RANSAC-like algorithms were developed for the reconstruction of $(d,{}^{2}\text{He})$ events \cite{ZAMORA2021164899}. In these routines, the  process of model verification is optimized by using a probabilistic approach. For instance, the candidate trajectories (model hypotheses) are evaluated on basis of an error probability distribution. This reduces the outlier sensitivity of \textsc{ransac} and provides a more robust method to identify particle tracks in point clouds with a large fraction of outliers. Also, a modification in the sampling process based on the relative distance and deposited charge was included to preferentially generate more useful \textsc{ransac} hypotheses. The random sampling is constructed in
a way that neighboring points are selected with higher probability by assuming a Gaussian distribution for the relative distance between points. Similarly, a probability density distribution using the total charge of the point cloud allows to select more efficiently the  points that have a larger energy deposited in gas \cite{ZAMORA2021164899}.  Finally, the tracking algorithm is coupled to a least squares routine that improves the  \textsc{ransac} output and provides a better estimate of the reaction vertex and scattering angles. \par

For $(d,{}^{2}\text{He})$ experiments, the AT-TPC operates without a magnetic field. Thus, the particle  tracks are 3D line trajectories. The tracking algorithm fits all particle tracks in each event and identifies $(d,{}^{2}\text{He})$ reactions as two protons emitted with the same vertex (see  Fig.~\ref{attpc}). The momenta of the recoiling protons ($\bm{P}_{px}$ with $x=1,2$) are extracted by combining the  track fitting result and the respective track length. The proton energy in the gas can be directly extracted from stopping power tables calculated with the code \textsc{srim} \cite{ZIEGLER20101818}. This information is sufficient for the reconstruction of a 
${}^{2}\text{He}$ particle. The first step is to obtain the momentum of ${}^{2}\text{He}$ from the proton tracks
\begin{equation}
 \bm{P}_{{}^2{\text{He}}} = \bm{P}_{p1} +\bm{P}_{p2}.
\end{equation}
Then, the relative energy between the two protons is extracted from the invariant-mass calculation in the ${}^{2}\text{He}$ frame
\begin{equation}
 \varepsilon_{pp} = \sqrt{E_{{}^2\text{He}}^2 - \bm{P}^2_{{}^2{\text{He}}}} - 2m_p,
\end{equation}
where $E_{{}^2\text{He}}$ is the total energy of the ${}^{2}\text{He}$ particle and $m_p$ the proton mass.

\begin{figure*}[!ht]
\centering
\includegraphics[width=0.4\textwidth]{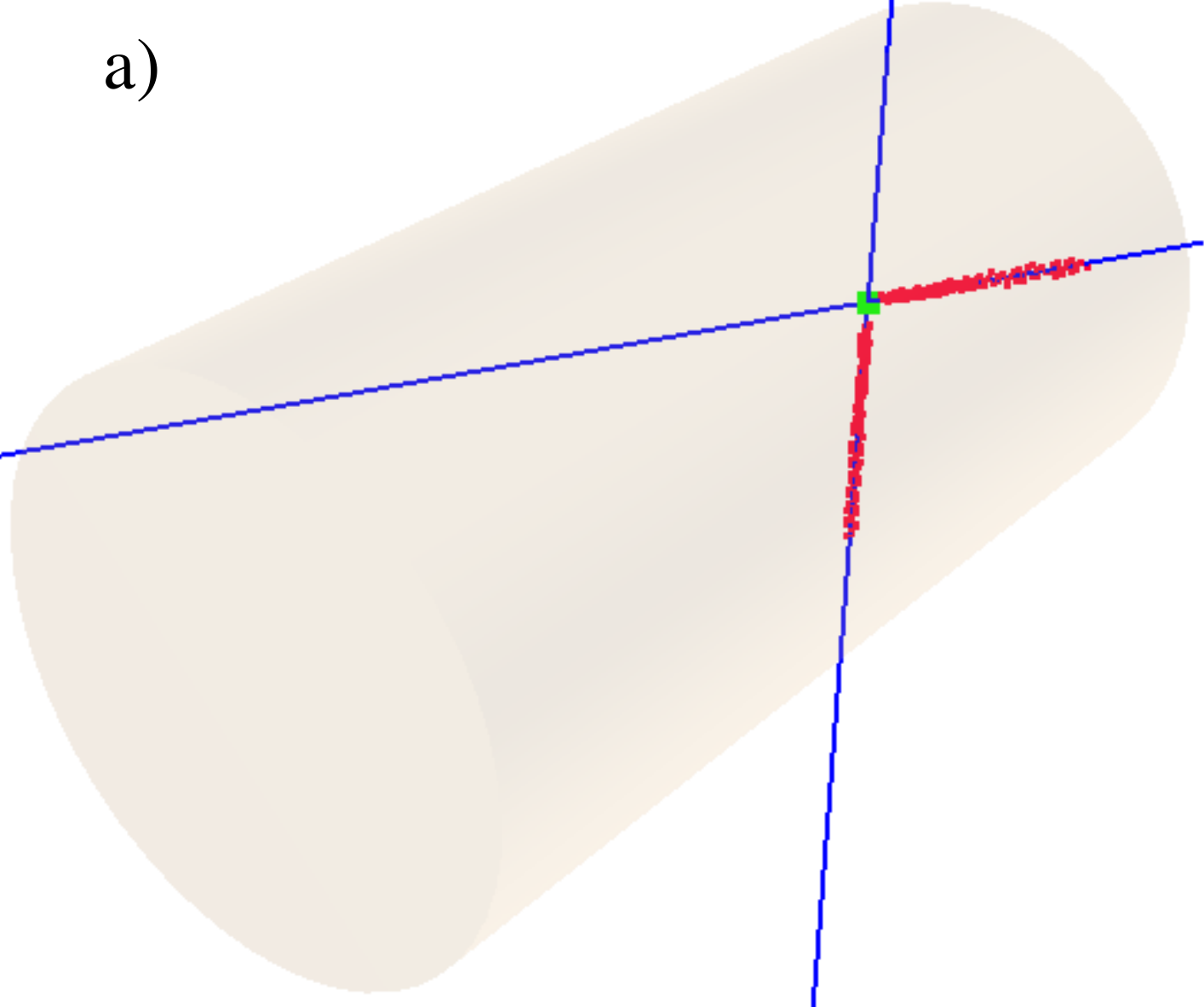}
\includegraphics[width=0.45\textwidth]{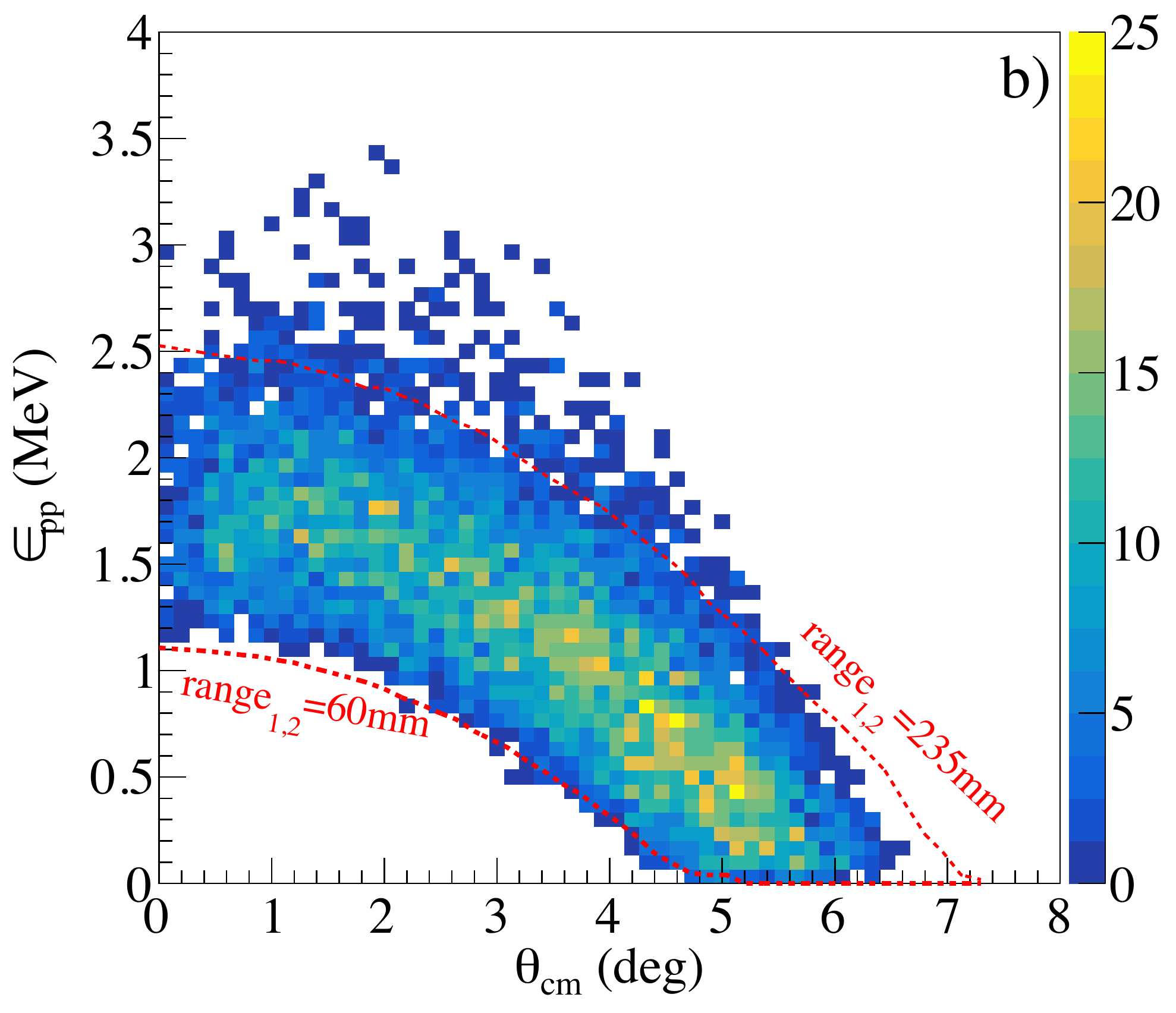}\\
\includegraphics[width=0.45\textwidth]{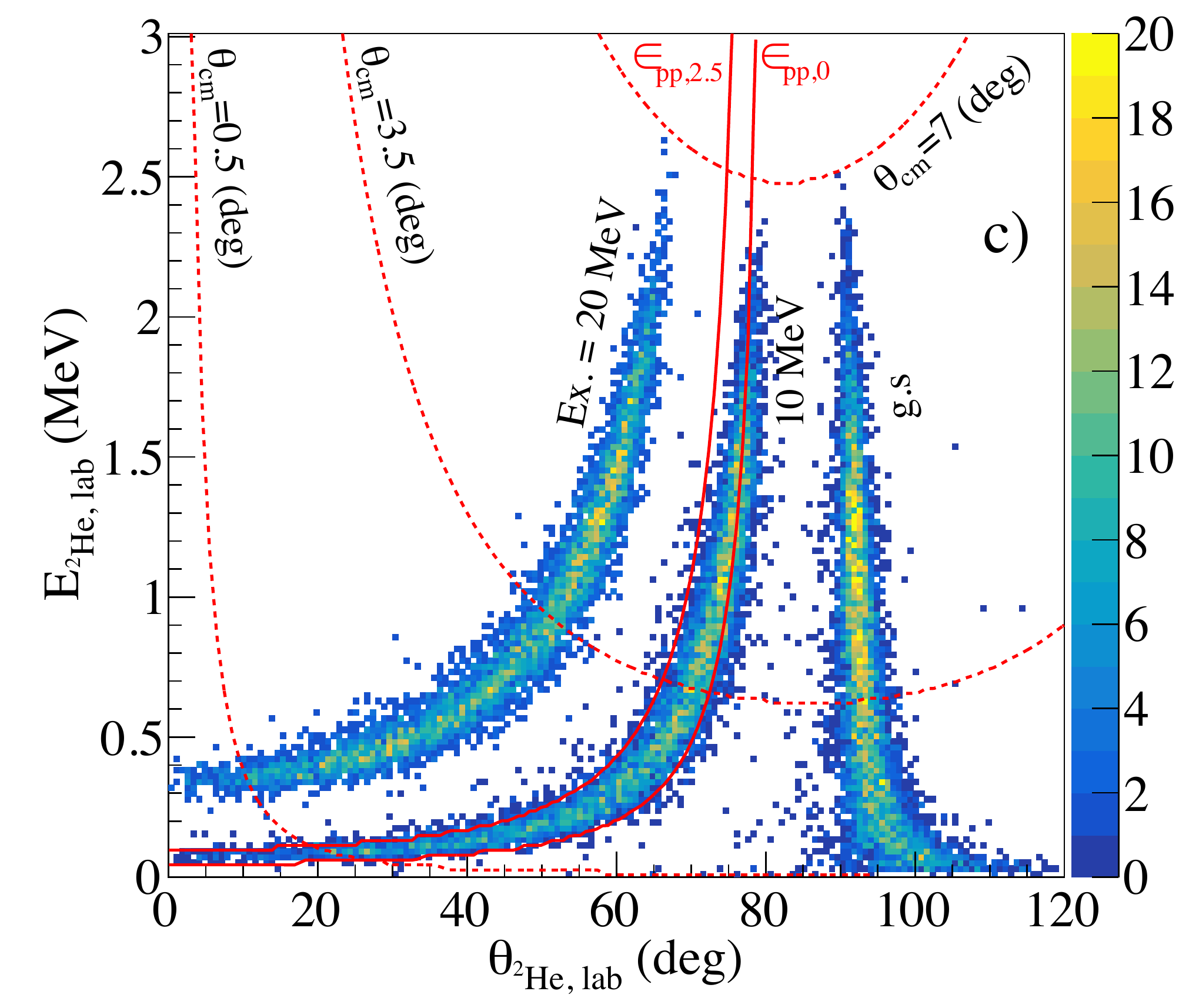}
\includegraphics[width=0.43\textwidth]{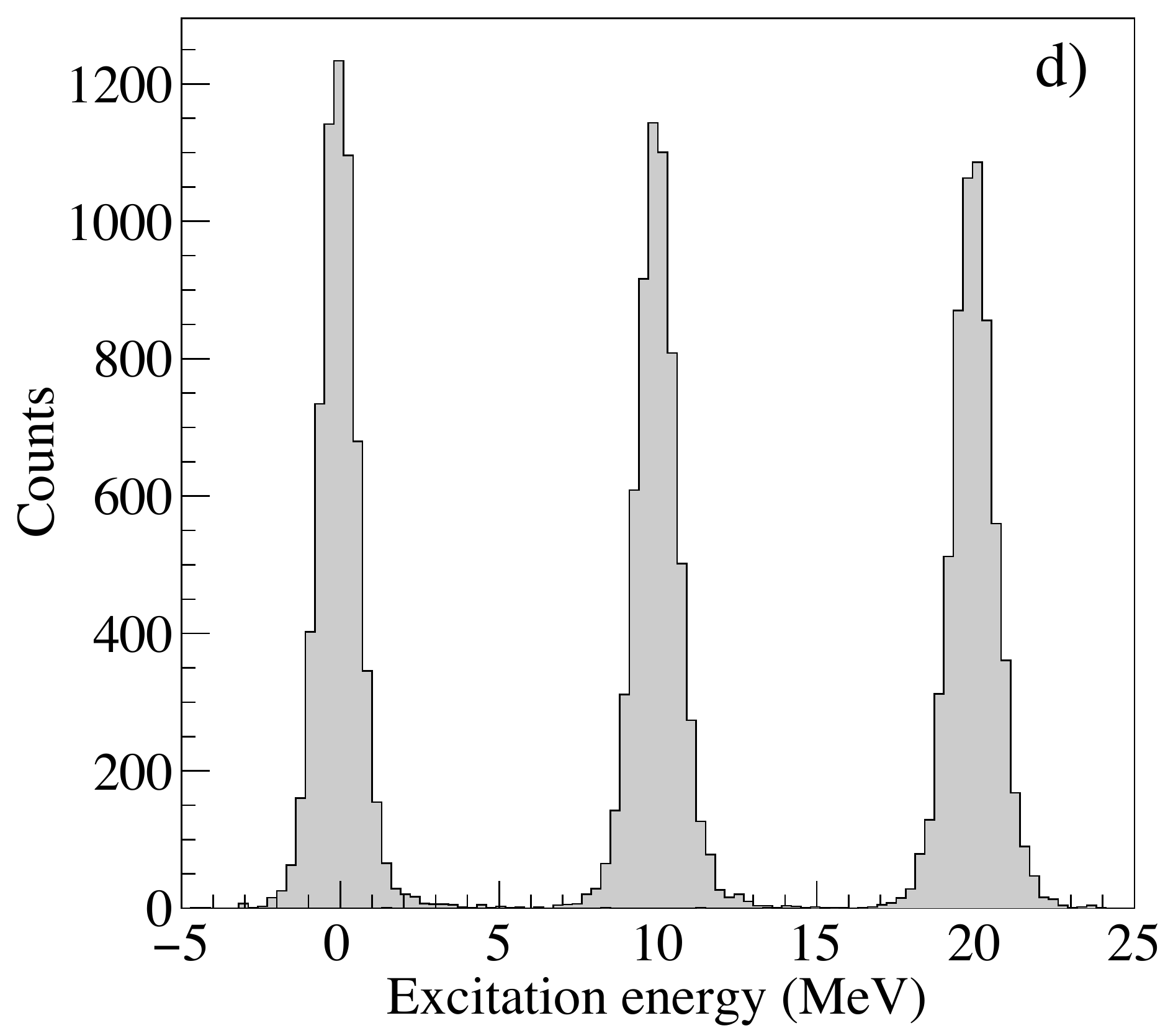}
\caption{\label{kine} a) Simulated $^{14}$O(d,$^2$He)${}^{14}$N event, at a beam energy of 100~MeV/u,  with the respective track fitting and vertex reconstruction. b) Internal energy of $^2$He as function of the center-of-mass scattering angle. The red dashed lines indicate the particle ranges of 60~mm and 235~mm, respectively.   The points above the 235~mm range correspond to tracks with scattering angles different than 90\textdegree.  c) Kinematics for the $^{14}$O(d,$^2$He)${}^{14}$N reaction, reconstructed $^2$He scattering angle as function of $^2$He kinetic energy. Three  excited states in $^{14}$N were assumed in this example: E$_{x}$= 0, 10 and 20~MeV.  The red lines are theoretical kinematics lines at 10~MeV excitation energy for  $\varepsilon_{pp}$ values of 0 and 2.5~MeV. The black dashed lines indicate center-of-mass scattering angles (0.5\textdegree, 3.5\textdegree and 7\textdegree).  d) Reconstructed missing mass (excitation energy) of $^{14}$N, for ground state, 10 and 20 MeV inputs. The distributions were normalized. The  peak at 10~MeV has a full width at half maximum (FWHM)  of about $ 1.6$~MeV.}
\end{figure*}

To ensure that two protons are in the spin-singlet (${}^1\text{S}_0$) state \cite{Kox1993}, and the $(d,{}^{2}\text{He})$ reaction proceeds exclusively with the transfer of spin $\Delta S=1$, it is preferred to constrain the analysis to small values of $\varepsilon_{pp}$. In forward kinematics experiments, small values of  $\varepsilon_{pp}$ are selected through the limited acceptance of the spectrometers used to detect the two protons or through software cuts in the measured $\varepsilon_{pp}$ distribution \cite{PhysRevC.47.648,Rakers2002-2,Frekers2004}. However, in inverse kinematics, the integrated  $\varepsilon_{pp}$ range depends on the detection efficiency and acceptance of the detector system. At small $\theta_{\text{c.m.}}$ the momentum transfer of the reaction is also small, resulting in a low kinetic energy of the virtual ${}^{2}\text{He}$ particle. Consequently, the two protons from the decay of ${}^{2}\text{He}$ will also have small kinetic energies.  The reconstruction of the ${}^{2}\text{He}$ particles is limited by the size of the insensitive region of the AT-TPC (see Fig.~\ref{tracks}). In this case, the two proton tracks are detected when the relative energy $\varepsilon_{pp}$ is sufficiently high. This is shown in Fig.~\ref{kine}(b): relative energies between 1 and 2.5~MeV are probed for $\theta_{\text{c.m.}} \approx 0^\circ$. At larger $\theta_{\text{c.m.}}$, the momentum transfer increases, which is reflected in larger kinetic energies for the ${}^{2}\text{He}$ particle (also for the two protons). If the energy of the two protons becomes too high, they escape from the active volume of the AT-TPC and the reconstruction of the event is not possible. Therefore, at high $\theta_{\text{c.m.}}$, only events with $\varepsilon_{pp}\lesssim 1$~MeV can be reconstructed, as shown in Fig.~\ref{kine}(b). As the differential cross section for a particular transition depends on both $\theta_{\text{c.m.}}$ and $\varepsilon_{pp}$ (see Fig.~\ref{dcs}), a $\varepsilon_{pp}$ acceptance correction as a function of $\theta_{\text{c.m.}}$ must be taken into account for extracting the differential cross section and the GT transition strength. The simulations presented here are essential for this correction.   \par

Once the ${}^{2}\text{He}$ particle is reconstructed, a missing-mass calculation is performed to extract the excited states populated by in the reaction. Fig.~\ref{kine}(c) shows the reconstructed  kinematic plot for the ${}^{14}\text{O}(d,{}^{2}\text{He}){}^{14}\text{N}$ reaction ($Q_{\text{g.s.}} = 3.7$~MeV) and its respective projection in the missing mass (Fig.~\ref{kine}(d)). Simulations of $(d,{}^{2}\text{He})$ reactions populating the ground state (g.s.) and excitation energies of 10 and 20~MeV are assumed in order to test the reconstruction routines.

The broadening of the kinematic lines is not only due to the energy and angle resolutions of the reconstruction, the distribution of the internal $^2$He energy also contributes to it (e.g., about $\sim 30\%$ in the g.s. Fig.~\ref{kine}(c)). The reconstructed scattering angle and kinetic energy of the $^2$He particle at  $E_x=10$~MeV (see Fig.~\ref{kine}(c)) seems to deviate from two-body kinematics calculation at $\theta_{cm}\gtrsim 6$\textdegree, this is explained by the limited acceptance at large scattering angles (Fig.~\ref{kine}(b)) and the resolution of the reconstruction. The resolution of the excitation-energy peaks (Fig.~\ref{kine}(d)) depends on the kinematics of the reaction. For instance, the resolution of the ground-state transition in the present example is dominated by the uncertainty in the reconstructed scattering angle of the $^2$He particle, whereas at higher excitation energies the resolution depends on both the kinetic energy and the scattering angle. The large acceptance of the AT-TPC provides  a good kinematical reconstruction  on  almost the entire angular range. Due to the relative good angular resolution (better than $1.5^\circ$ in $\theta_\text{lab}$) achieved by the tracking  algorithm, the excited states are easily separable up to about $8^\circ$ in the center-of-mass system with an energy resolution of about 1.6~MeV (FWHM).

\section{Coincidences with beam-like particles \label{kinfit}}
\begin{figure}[!ht]
\centering
\includegraphics[width=0.5\textwidth]{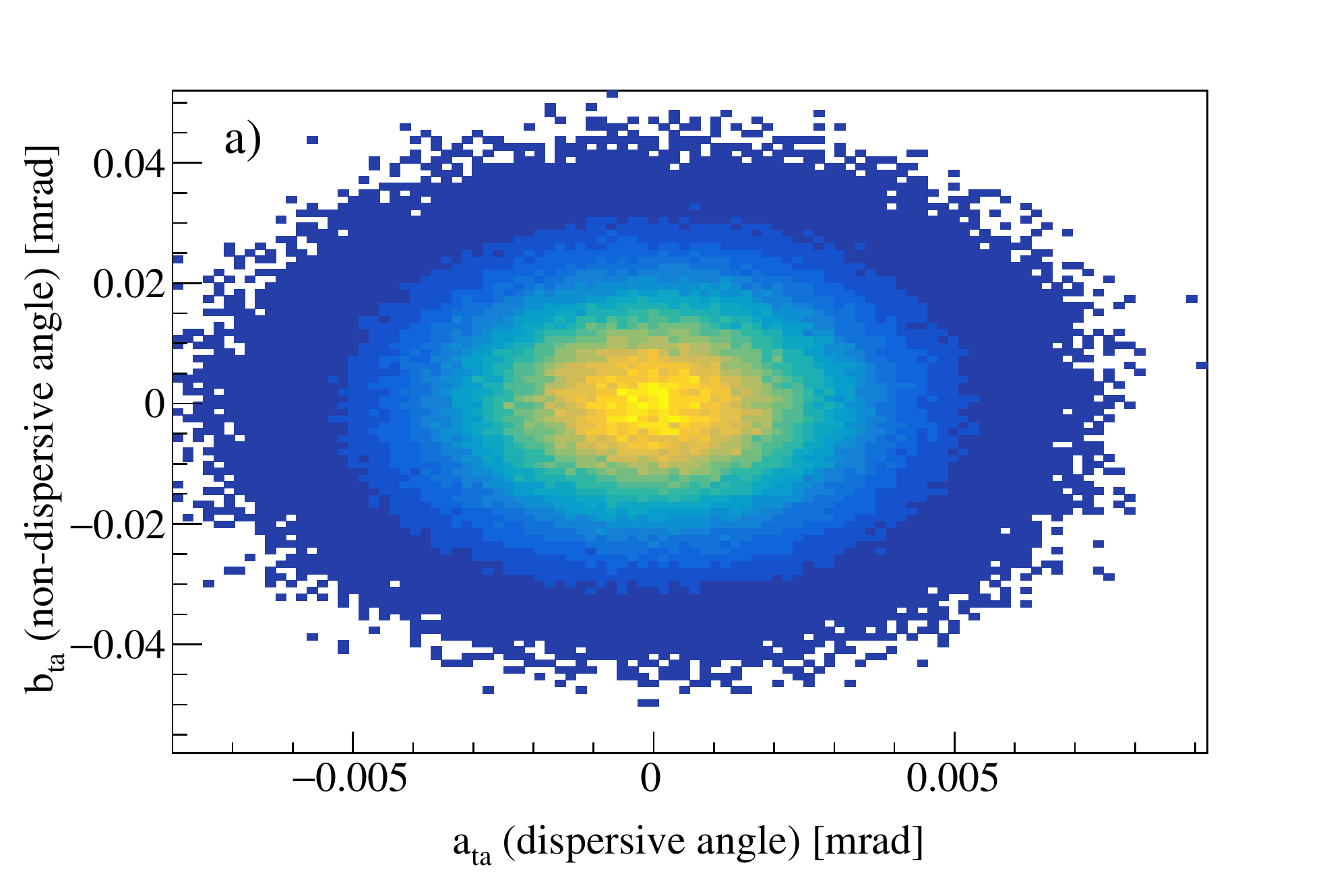}\\
\includegraphics[width=0.5\textwidth]{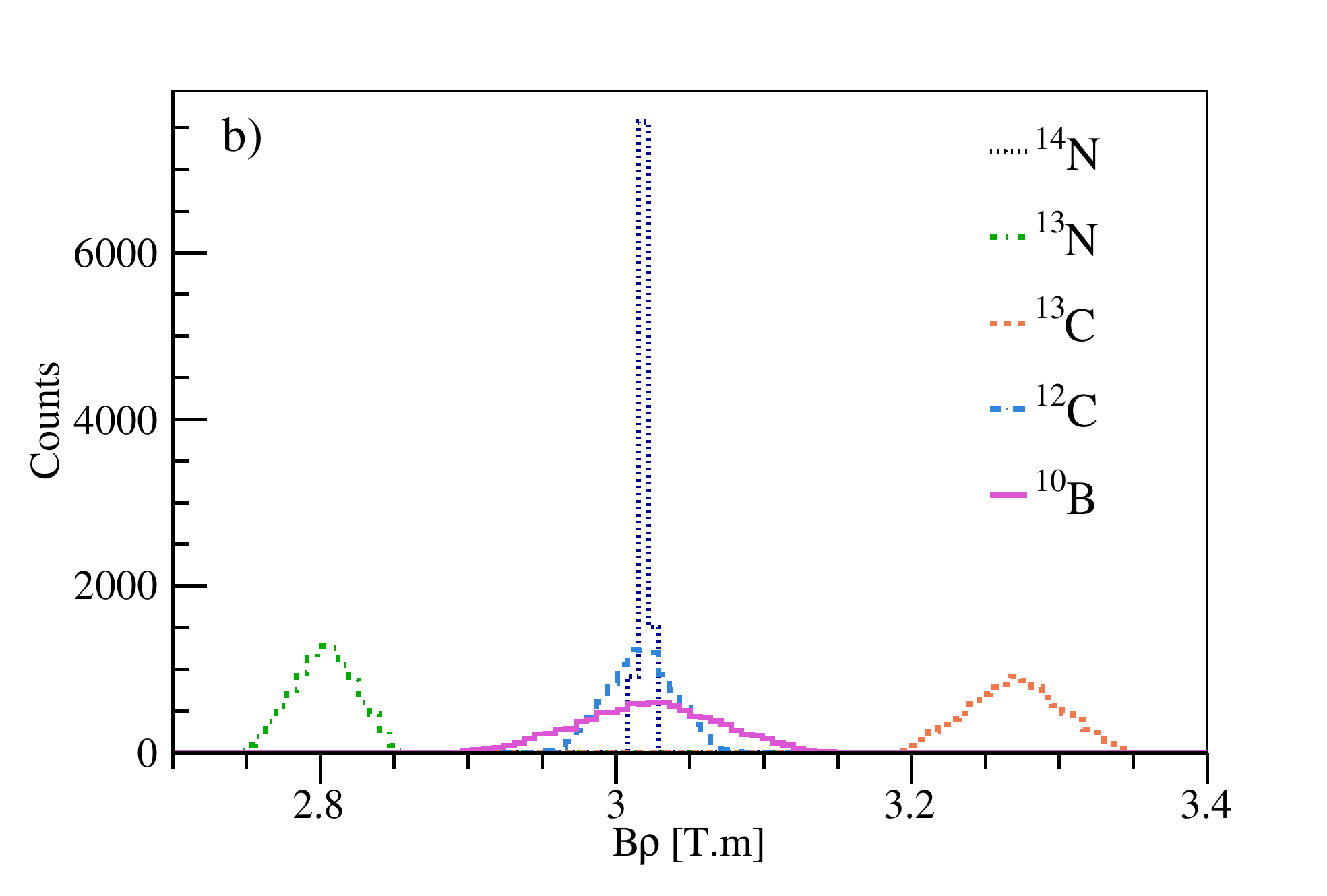}
\caption{\label{brho} (a) Simulated dispersive and non-dispersive angles of the ejectiles in the $^{14}$O(d,$^2$He)$^{14}$N reaction at the target location. (b) Magnetic rigidity of the ejectiles $^{14}$N and several decay products.}
\end{figure}

Complete kinematics measurements including the detection of  beam-like particles provide a  strong selectivity of the reaction mechanism and also reduce significantly the background in the experimental data. Therefore, a routine that includes a phase-space generator was implemented to properly simulate the in-flight decay and momentum of the residues. As the reaction ejectiles in inverse kinematics are boosted to forward angles in the laboratory system, the AT-TPC is coupled with the S800 spectrometer at FRIB \cite{BAZIN2003629}. Then, the $(d,{}^{2}\text{He})$ events can be correlated with the respective particle identification in the spectrometer focal plane detectors. For instance, Fig.~\ref{brho}(a) shows the simulated dispersive and non-dispersive angles of the beam-like particles accepted in the S800 spectrometer assuming an incident  ${}^{14}$O beam at 100~MeV/u. Also, from the momentum of the accepted residues it is possible to obtain the respective magnetic rigidity ($B\rho$). Fig.~\ref{brho}(b) shows the simulated $B\rho$ distributions for the  ${}^{14}\text{O}(d,{}^{2}\text{He}){}^{14}\text{N}$ reaction  including several decay channels such as $p$, $n$, $np$ and $\alpha$.\par

\begin{figure}[!ht]
\centering
\hspace*{-0.25cm}
\includegraphics[width=0.5\textwidth]{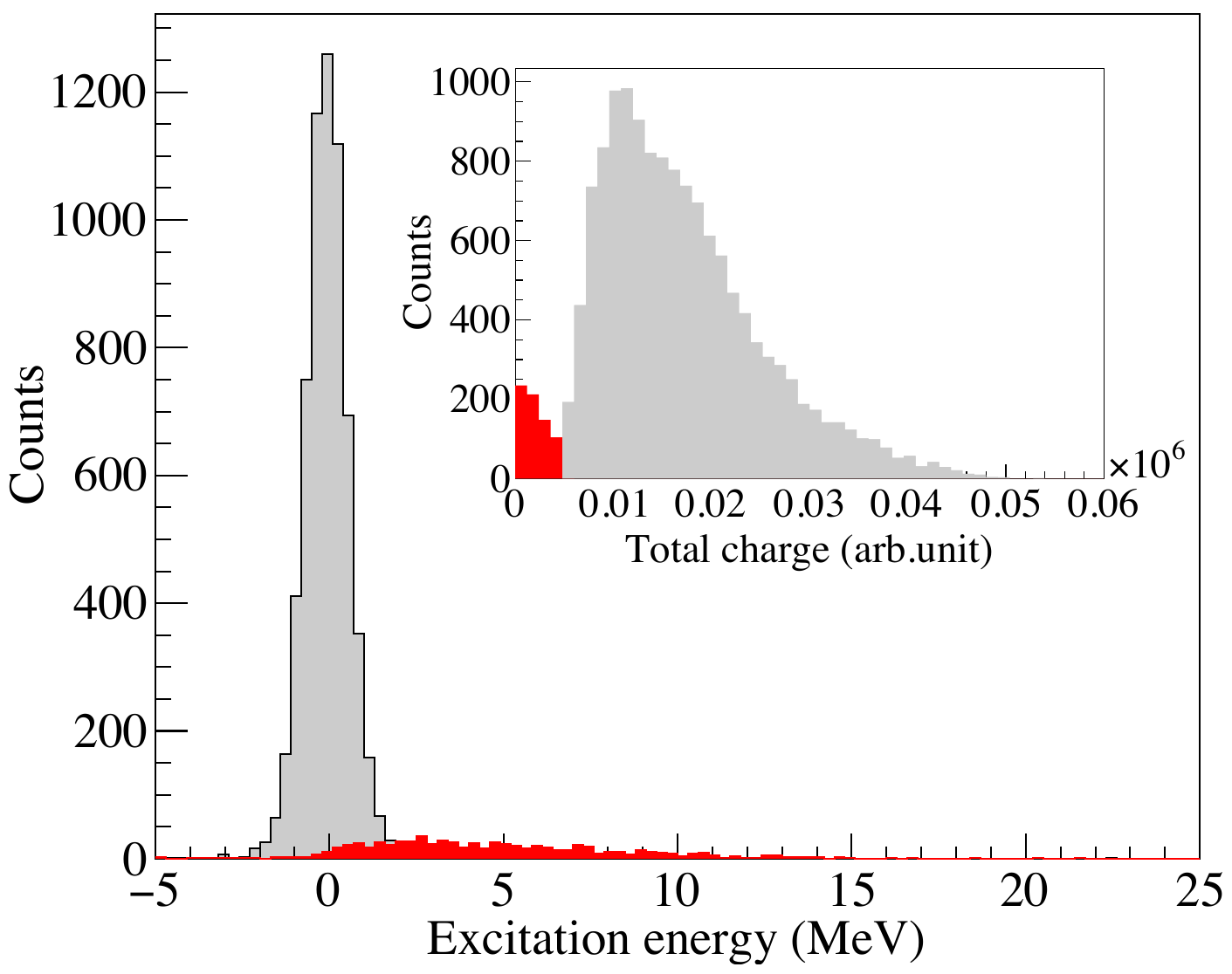}
\caption{\label{exbck} Reconstructed ${}^{14}$N ground-state energy distribution in the ($(d,{}^{2}\text{He}$) reaction (grey peak). The contribution due to stochastic noise fitting is shown in red. The inserted plot represents a distribution of the integrated charge per track and the small bump below $5\times 10^{3}$ (arb. unit) is associated to the red background in the excitation energy distribution.}
\end{figure}

The identification of a beam-like particle in coincidence with two protons emerging from the same vertex is a stringent condition that suppresses the background. As discussed before, $\delta$-rays are expected to be a significant contribution in the point clouds. These points are an important source of background that can lead to a wrong identification of $(d,{}^{2}\text{He})$ events. In this case, the integrated charge  and number of points of each track are substantially smaller than the ones expected for a real proton track.  Fig.~\ref{exbck} (insert) shows that these spurious tracks are easily suppressed by the integrated charge condition.  Random coincidences due to multiple-scattering events or  scattering close to a $(d,{}^{2}\text{He})$  reaction vertex can also be a source of background. However, the probability of occurrence for such events is very small, even for  beam rates at the level of $R= 10^6$ pps. For example, the probability to have an elastic scattering [${}^{14}$O$(d,d)$] event  during a time window of $W=100\times10^{-6}$ s (typical drift time of the electrons for a $L=100$~cm long drift volume) can be modeled by assuming a Poisson probability distribution with $\lambda = L \rho \sigma W R \approx 8.7\times10^{-3}$. In this case, $\rho \approx 1.74 \times 10^{19}$~cm$^{-3}$ is an estimate value of the density of atoms in the gas target, and the scattering cross section $\sigma \sim 50\times10^{-27}$ cm$^2$ ($\theta_\text{c.m.} \lesssim$ 8 deg). Using the Poisson distribution with this information, one finds that the probability to have at least one deuteron elastically scattered  within the same time window of a $(d,{}^{2}\text{He})$ reaction is in the order of 1\%. Therefore, the background contribution due to an accidental detection of two deuterons elastically scattered, with vertices within $\sim 1$~cm [similar condition as the two protons of a $(d,{}^{2}\text{He}$)  reaction], is reduced by a factor 100, which makes this source of background negligible.

\subsection{Kinematic fitting}
 In complete kinematic measurements, the resolution of the reconstructed quantities can be improved  by using a mathematical procedure called kinematic fitting (KF). KF is a widely used technique in high-energy physics \cite{kinfit}. Recently, this method has been proposed for the analysis of nuclear reactions at low energy \cite{physics1030027}. The goal of the kinematic fitting is to reduce the uncertainties from the measured quantities by minimizing the reaction kinematics using a few constraints such as the momentum and energy conservation. The solution method implemented in this work is based on the kinematic fitting programs \textsc{fit70} \cite{fit70} and \textsc{kwfit} \cite{kinfit}. \par
 
\begin{figure}[!ht]
\centering
\includegraphics[width=0.5\textwidth]{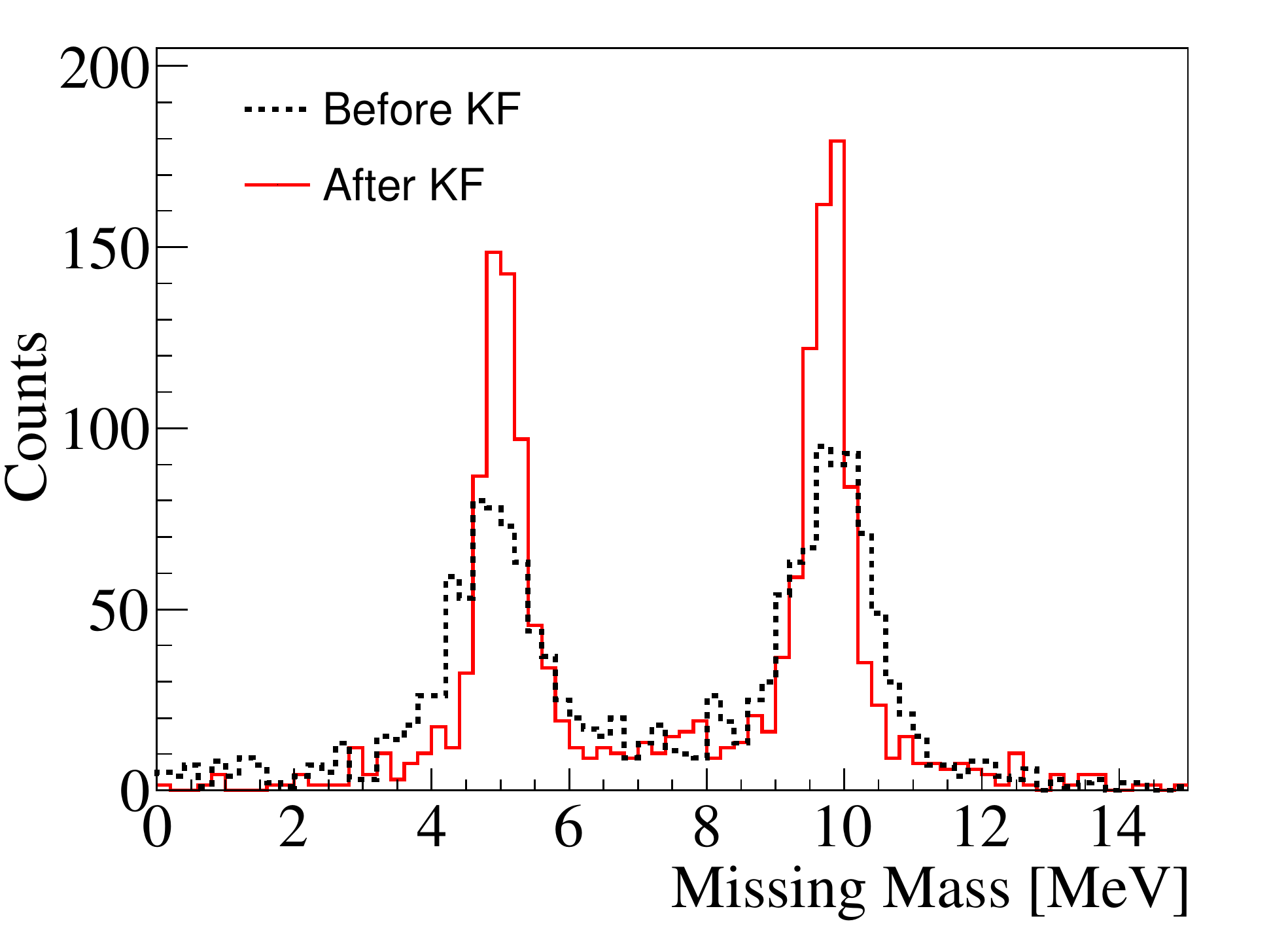}
\caption{\label{kf} Missing-mass reconstruction before (dashed line) and after (solid line) the kinematic fitting. Arbitrary states at 5 and 10~MeV excitation energy were assumed to demonstrate the performance of the technique. }
\end{figure}

 In the particular case of a $(d,{}^{2}\text{He})$ reaction, the input channel corresponds to a projectile and a deuterium target, while the output channel is composed by two protons and an ejectile. Each particle have a four-momentum  $\alpha_i = (p_{xi},p_{yi},p_{zi},E_i)$  vector that is extracted from the observables of an experiment. The deuteron target is at rest in the laboratory frame, and therefore, this  information is only included as a reaction constraint.  Let $\alpha$ represent the column vector that contains the four-momentum vectors of the projectile and the three particles in the output channel
 \begin{align}
    \alpha &= \begin{pmatrix}
           \alpha_{1} \\
           \alpha_{2} \\
           \alpha_{3} \\
           \alpha_{4}
         \end{pmatrix}
  \end{align}
The  constraints assumed in the KF are the momentum (3D) and the total energy conservation as
\begin{equation}
 {\bf H}(\alpha)\equiv 0, \hspace{0.3cm} \text{where} \hspace{0.3cm} {\bf H} = (H_1, H_2, H_3, H_4).
\end{equation}
The equation of above can be expanded around a convenient point $\alpha_A$ that produces a set of linear equations \cite{kinfit}
\begin{equation}
 0= \frac{\partial {\bf H}(\alpha_A)}{\partial \alpha}(\alpha -\alpha_A)+{\bf H}(\alpha_A)= {\bf D}\delta \alpha +{\bf d},
\end{equation}
where ${\bf D}$ is a matrix of dimension $4\times 4$ that contains the partial derivatives of the constraint parameters $D_{ij} = \frac{\partial Hi}{\partial \alpha_j}$,  and ${\bf d}$ is a vector of constraints evaluated at the point $\alpha_A$.  The fitting technique is based on the Lagrange multipliers method, in which the $\chi^2$, given by
\begin{equation}
 \chi^2 = (\alpha-\alpha_0)^TV_{\alpha0}^{-1}(\alpha-\alpha_0)+2\lambda^T({\bf D}\delta \alpha +{\bf d}), \label{chi2}
\end{equation}
is minimized. Here, $V_{\alpha0}$ is the covariance matrix and $\lambda$ is a vector of Lagrange multipliers. A routine for minimizing Eq.~(\ref{chi2}) was implemented in the code. The resulting parameters $\alpha$ are improved  (compared to the input values $\alpha_0$) by the momentum and energy constraints of the reaction. In order to avoid problems with the reconstruction of relative energy $\varepsilon_ {pp} $, an anticorrelation between the two protons in the output channel was defined in the covariance matrix. The KF was tested with the reconstruction of two peaks at 5 and 10~MeV, as shown in Fig.~\ref{kf}. As the beam particles interact with the gas in the insensitive region, no information of the beam momentum is possible to be extracted from the point clouds. For the KF,  it was assumed the beam particle impinging the AT-TPC in the normal direction with a momentum resolution (defined in the covariance matrix) of 0.1\%. The momentum of the ejectile particle was extracted directly from the event generator with the respective acceptance provided by the S800 spectrometer, as shown in Fig.~\ref{brho}. As explained in Section~\ref{reco}, the energy resolution achieved from the tracking is about 1.6~MeV, but with the kinematic fitting  the resolution is improved to a value of 0.8~MeV. Thus, the minimization routine corrects for the uncertainty in the energy and angles of the proton tracks by using energy and momentum conservation. Therefore,  a significant improvement in the reconstruction of the invariant mass is expected for complete kinematic measurements. The KF routine presented here can also be implemented  in the analysis of future experiments.  In this case, it might be required the use of tracking detectors before the AT-TPC in order to extract with a better precision the projectile momentum and reaction vertex. Also, it would be necessary to investigate in more detail the reconstruction of virtual particles emerging from the decay of the residues of  $(d,{}^{2}\text{He})$ reactions.

\section{Space-charge effects \label{spacechar}}
\subsection{Charge build-up \label{subsec:chargeBuildUp}}

\begin{figure*}[!ht]
\centering
\includegraphics[width=0.45\textwidth]{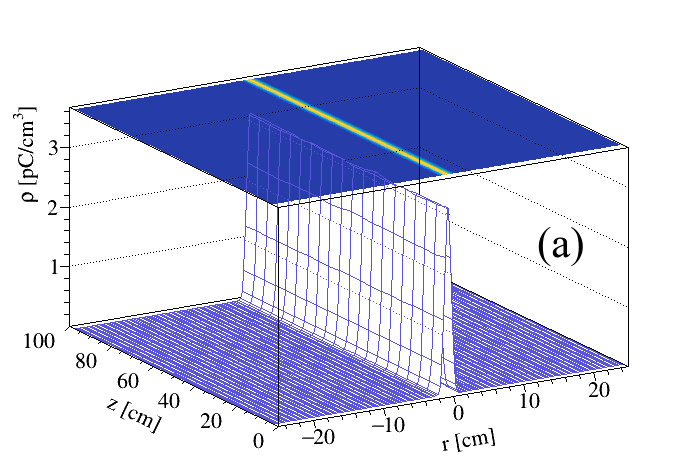}
\includegraphics[width=0.45\textwidth]{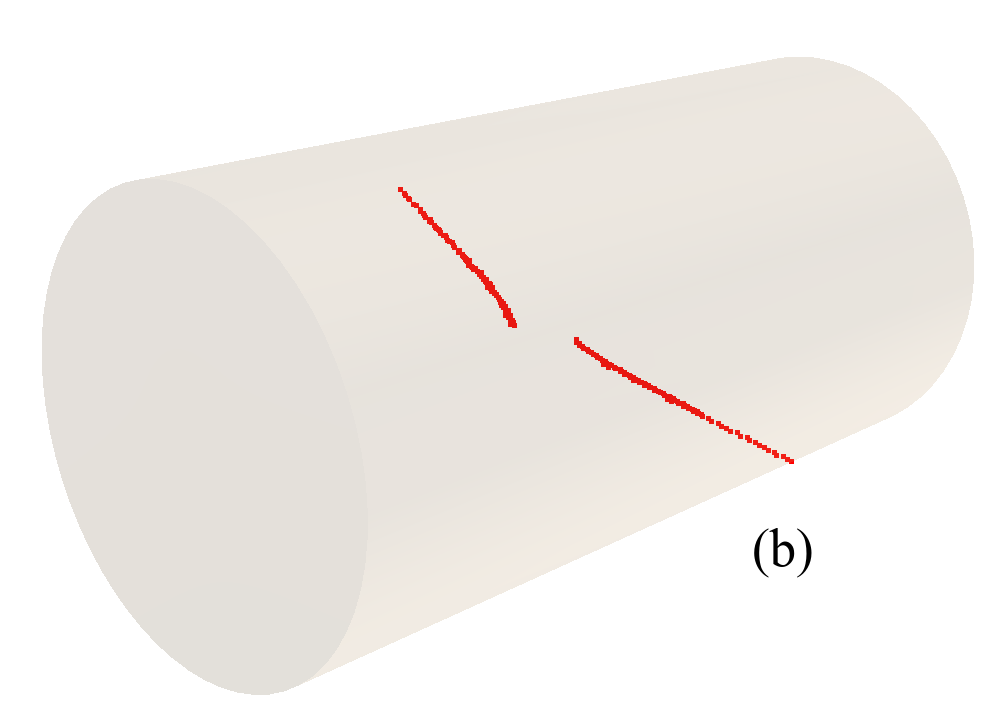}
\includegraphics[width=0.45\textwidth]{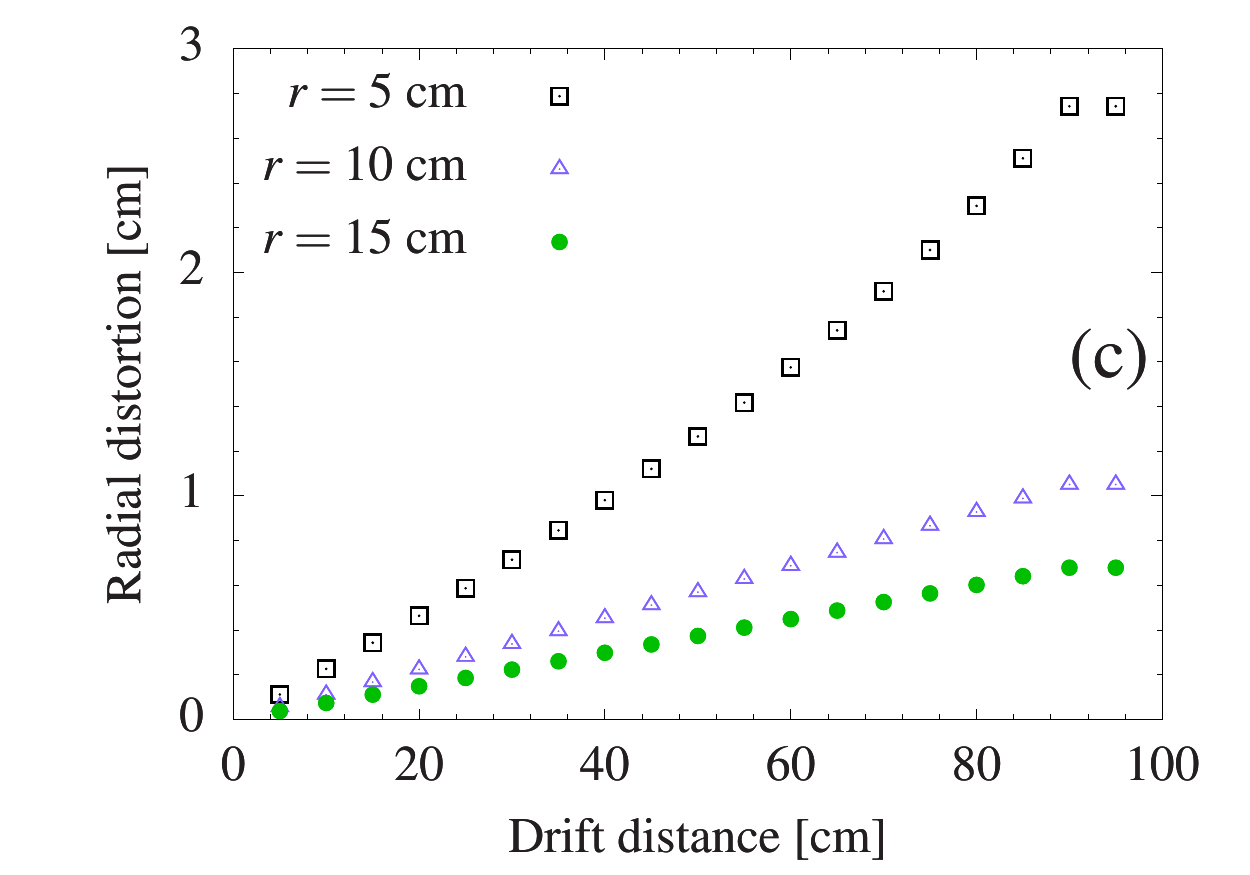}
\includegraphics[width=0.45\textwidth]{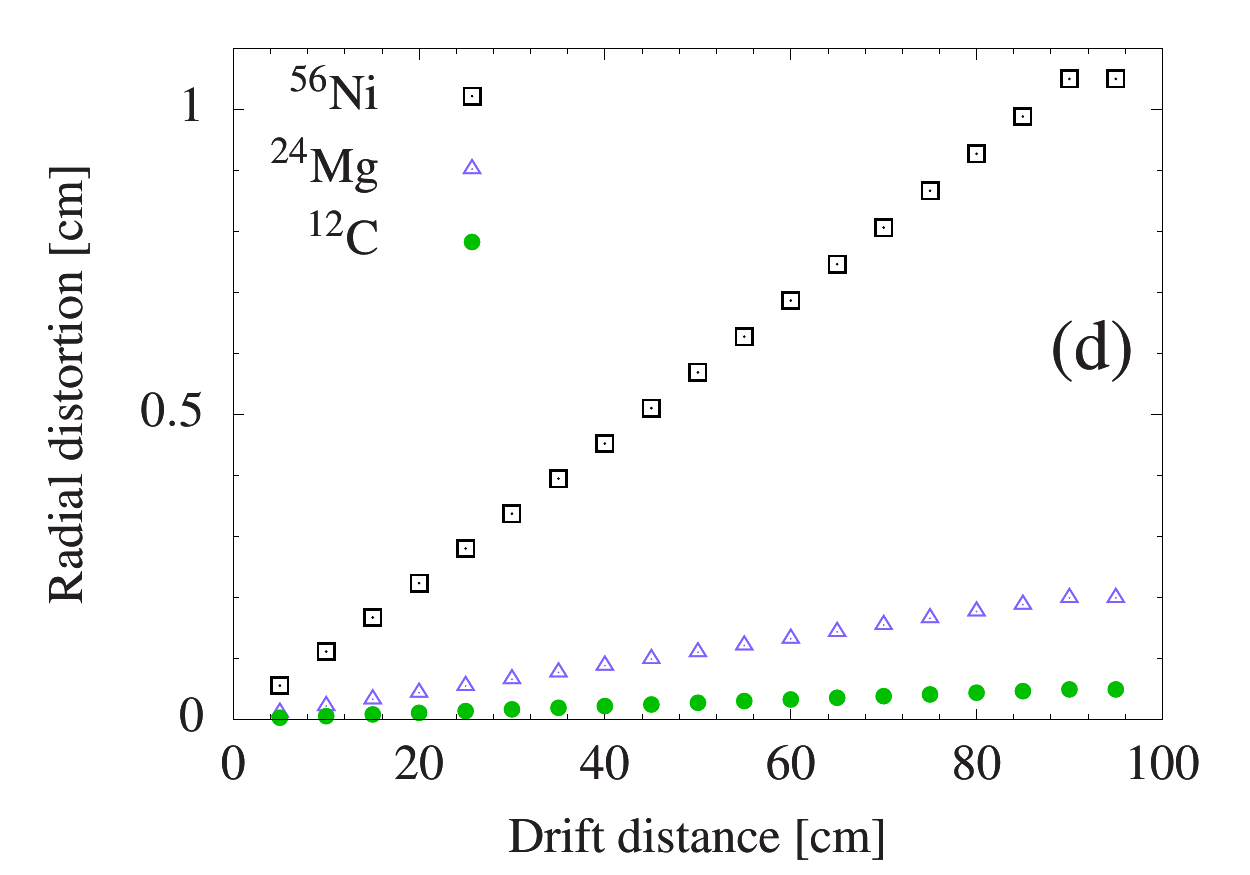}
\caption{\label{spch} (a) Charge distribution assumed in the model to describe the ion build-up. The distribution is a Gaussian profile extended along the active volume. (b) Simulated ($d$,$^{2}$He) event including space-charge effects.  An electric field component in the radial direction distort the tracks close to the beam axis. (c) Radial distortion as a function of the drift distance. The effects are stronger close to the beam axis. (d) Radial distortion as a function of the drift distance for three ion beams. The distortion effect increase with the ion-beam charge. }
\end{figure*}

Space-charge effects due to ion accumulation inside the AT-TPC are expected to be present at high beam intensities. A large number of electron-ion pairs [$N(r,z)$] are produced in the gas volume along the beam path. While electrons drift relatively fast (tens of $\sim\mu$s) to the pad plane, the drift time of ions is  a few  orders of magnitude slower. It means that the beam is continuously ionizing the gas target producing a charge build-up in the central region of the AT-TPC. This accumulated charge can produce a distortion of the electric field and also affect the electron drift. Experimental evidences of such space-charge effects in the AT-TPC have recently reported \cite{RANDHAWA2019162830}. Assuming a constant beam rate $R$, the production rate of electron-ion pairs per unit volume can be roughly approximated by
\begin{equation}
\label{ionrate}
 \dot{N} = \left( \frac{dE}{dx}\right) \frac{R}{wA},
\end{equation}
where $\left( \frac{dE}{dx}\right)$ is the beam stopping power in the gas, $A$ is the beam spot area and $w=37$~eV  is the mean ionization energy for $D_2$  \cite{HUYSE2002535}. The charge density in the AT-TPC volume at a time $t$ is given by \cite{BOHMER2013101}
\begin{equation}
 \rho(x,y,z,t) =  \int_0^t e \dot{N}(x,y,z,t')dt',
\end{equation}
with $e$ the electric charge unit. As an example,  a  ${}^{56}$Ni beam at 100~MeV/u and rate of $10^5$~pps interacting with the AT-TPC filled with a deuterium gas at 500~Torr are assumed. A realistic beam profile of a Gaussian distribution of $\sigma= 0.5$~cm is assumed. The  accumulated charge density after an integration time of $t_\text{max} = L/u^+$ ($L$ is the AT-TPC length and $u^+$ is the ion drift velocity) can yield up to 3~pC/cm$^3$ around the beam axis, as is shown in Fig.~\ref{spch}(a). In order to have a rough estimation of the electric field produced by  this charge density, a radial field component is calculated using Gauss' law in the approximation of an infinite cylinder  
\begin{equation}
 E(r) = \frac{e}{\varepsilon_0}\left( \frac{dE}{dx}\right) \frac{R t_\text{max}}{w} \left(\frac{1-\exp\left(-\frac{r^2}{2\sigma^2} \right)}{r}\right),
\end{equation}
where $\varepsilon_0$ is the permittivity of free space. A more rigorous solution can be obtained by integrating the corresponding Green's function, as  explained in Ref.~\cite{ROSSEGGER201152}. The resulting effect of this field component is a distortion of the electron drift distance. In order to evaluate such effects, it is necessary to simulate the electron transport by solving the Langevin equation numerically \cite{BOHMER2013101,ROSSEGGER201152}. An electron transport routine was implemented in the code in the pre-digitization module. An example of a  $(d,{}^{2}\text{He})$ reaction in a distorted electric field is shown in Fig.~\ref{spch}(b). As can be noticed, the tracks are mostly affected in the central region of the AT-TPC. These space-charge effects may cause problems in the event reconstruction, in particular, for the extraction of scattering angles and reaction vertex. The radial distortion is defined as the relative distance between the projected initial point of the electron $(x_0,y_0)$ and the final point on the pad plane $(x_f,y_f)$. Depending on the drift distance, the radial distortion can increase because of the electron transport. For instance, one can see the distortion at different initial distances relative to the beam axis $(r)$ in Fig.~\ref{spch}(c). The space-charge effects are stronger near the beam axis, and they become negligible for electrons produced close to the wall. For an initial distance of $r=5$~cm, the distortion can be as big as 3~cm, after a drift distance of 100~cm, which is approximately the equivalent distance of 6 pads. The space-charge effects have also a significant dependence on the beam particle due to the amount of electrons produced in the beam energy loss. The higher the beam charge, the stronger the radial distortion is. Fig.~\ref{spch}(d) shows an example of the radial distortion at  $r=10$~cm for different beams.  As expected, the space-charge effects are higher for ${}^{56}$Ni which are about a factor 4 stronger than for ${}^{24}$Mg.

\subsection{Recombination \label{subsec:Recombination}}
Electron recombination is also an important space-charge effect that needs to be taken into account in the simulation. For this effect, the electrons produced by ionizing particles in the gas volume may recombine with nearby ions due to electromagnetic interaction. For high-ionization densities, the recombination effect can induce significant losses in the collected charge. Two types of recombination processes are possible in the AT-TPC, columnar or volume recombination \cite{HUYSE2002535}. The columnar process is when an ion-electron pair produced in the same track recombines, while the volume process is a more general case where the ions are recombined with electrons from different tracks. Due to the relative strong fields and highly ionizing particles used in the AT-TPC, the columnar recombination is expected to be negligible in comparison with the volume recombination \cite{RANDHAWA2019162830,HUYSE2002535}. In the latter case, the recombination depends on the ionization rate $\dot{N}$, which can be approximated by Eq.~(\ref{ionrate}). Thus, the recombination effect is beam-rate dependent. The time scale required for an ion-electron pair to recombine depends on the intrinsic properties of the gas expressed by the recombination coefficient $\alpha$, which for a deuterium gas is $\alpha \sim 2.5\times 10^{-6}$~cm$^3$/s \cite{recoeff}.\par
The recombination loss  in  ionization chambers (parallel plate chambers) can be estimated as \cite{COLMENARES1974269}
\begin{equation}
\label{recomb}
    f = \frac{\alpha \dot{N} L^2}{6 \mu^2 E^2},
\end{equation}
where $\mu$ is the ion mobility and $E$ is the strength of the electric field. Eq.~\ref{recomb} provides an estimate value for the loss in charge collection assuming a constant beam rate impinging the gas target. As an example, Table~\ref{tablerecob} shows the recombination loss for three beams at 100~MeV/u and $10^5$~pps interacting with a deuterium gas at 500~Torr and field strength of 500~V/cm.

\begin{table}[htb!]
\caption{Recombination loss and electron-ion production rate in a deuterium gas at 500~Torr and 500~V/cm.  Three  different beams were assumed at 100~MeV/u and $10^5$~pps. }
\setlength{\tabcolsep}{20pt} 
\renewcommand{\arraystretch}{1.7} 
\centering
\begin{tabular}{ccc}
\hline\hline
Ion Beam              & $\dot{N}$ [cm$^3$/s]   & $f$ [\%] \\ \hline
            
 $^{12}$C   &    $5.5\times 10^{7}$    &    0.5     \\
 $^{24}$Mg   &    $2.1\times 10^{8}$    &    1.7     \\
 $^{56}$Ni   &    $1.1\times 10^{9}$    &    9.1     \\
\hline \hline
\end{tabular}
\label{tablerecob}
\end{table}

As can be noticed, the recombination increases with the beam energy loss in the gas. As the recombination is expected to occur most likely at near  the beam region, the absolute effect may be  a reduction in the space-charge build-up that distorts the electric field. For beam intensities above $10^6$~pps,  the recombination effect becomes significant and it is expected that part of the information of the tracks is lost \cite{RANDHAWA2019162830}.

\subsection{Correction}
\label{subsec:Correction}
\begin{figure}[!ht]
\centering
\includegraphics[width=0.5\textwidth]{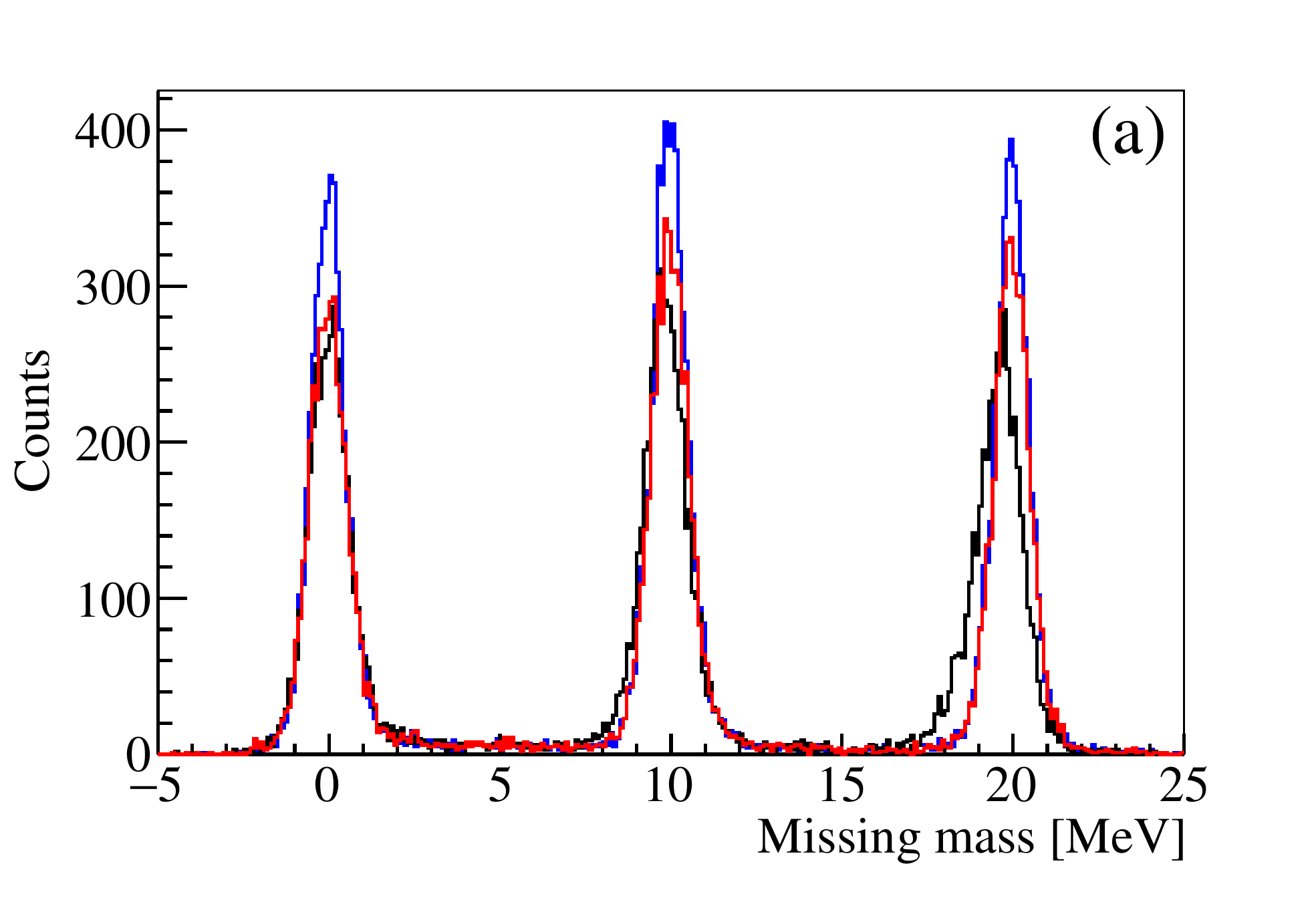}
\includegraphics[width=0.5\textwidth]{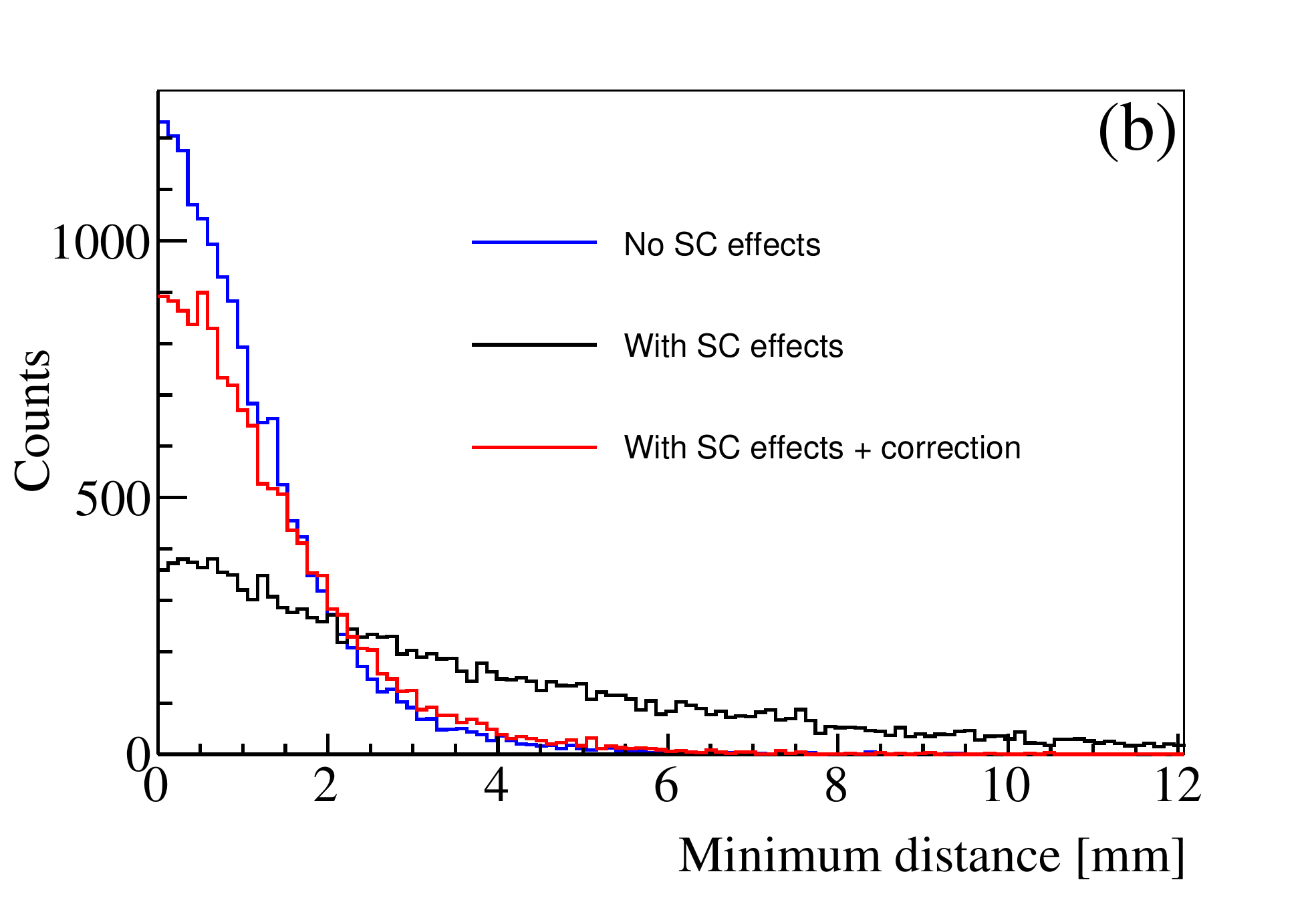}
\caption{\label{schcorr} Simulated $^{14}$O($d,^{2}$He) data with and without space-charge effects and the respective correction using a distortion map. (a) Reconstructed missing-mass energy. (b) Distribution of the  minimum distance between the proton tracks.  }
\end{figure}

The space-charge effects, mentioned in \ref{subsec:chargeBuildUp} and \ref{subsec:Recombination} can be corrected for using a  distortion map of the full active volume generated from simulations. As the beam impinges in central region of the active volume, space-charge effects can be assumed to be axially symmetric. This reduces the extraction of a distortion map to a two-dimensional problem. An electron transport routine including charge build-up and recombination effects was used to generate and transport electrons at different initial positions along the whole active volume. The relative distance of the electron cloud on the pad plane with respect to a given initial electron position defines the distortion. The distortion map was applied directly to the geometrical coordinates of each hit in order to restore the  particle tracks.  Fig.~\ref{schcorr}(a) shows a comparison of the reconstructed missing-mass energy for simulations with and without space-charge effects, and after the correction based on the distortion map. As can be seen, the energy resolution is strongly affected by the space charge mostly at high excitation energies. This effect can be related with a wrong determination of the reaction vertex and scattering angle of the protons. Similarly, the reconstruction efficiency of the $^2$He particle is reduced by about 13\% when the space-charge effects are included,  i.e., 87\% of the tracks can be restored with the present correction.  This reduction in efficiency is because a few short tracks lose a fraction of their points in the central dead region (pad plane hole). However, with the correction, one is able to restore the energy resolution of the peaks. The minimum distance between the fitted proton tracks in the central region gives an estimate of how well the vertex can be reconstructed, see Fig.~\ref{schcorr}(b). The space-charge effects produce a  wider distribution of the minimum distance between the two proton tracks that leads to a larger uncertainty in the determination of the vertex point. For this study, we chose the minimum distance between tracks to be less than 1~cm, which corresponds to about $2\sigma$ of the distorted track distribution. The correction improves the vertex reconstruction, which also impacts the missing-mass energy, as shown in Fig.~\ref{schcorr}(a). A similar procedure can be employed to correct experimental data. In this case, the  distortion map might be extracted from simulations or from the same data by quantifying the distortion of the  tracks in the full active volume.

\section{Summary}
A simulation and analysis code has been developed to study charge-exchange $(d,{}^{2}\text{He})$ reactions in inverse kinematics using the AT-TPC. The simulation package provides a realistic detector response that allows for processing the simulation output in the same manner as the experimental data. Dedicated subroutines for the event generator, digitization and reconstruction of nuclear reactions are used for simulating experiments with the AT-TPC. In particular, the event generator of $(d,{}^{2}\text{He})$ reactions takes into account the production of two recoiling protons at low-momentum transfer with a relative energy,  based on adiabatic coupled-channel calculations that are known to reproduce the dependence of the $(d,{}^{2}\text{He})$ differential cross section on $\varepsilon_{pp}$ and $\theta_{cm}$. A point cloud of the reaction is obtained from the digitization of the proton tracks in the active gas volume. Pattern recognition and fitting algorithms are used for the analysis of the point clouds and for reconstructing the $(d,{}^{2}\text{He})$ reactions. The missing-mass energy is very well  reconstructed independently of the geometrical acceptance of the detector.\par

A few analysis tools were successfully tested with simulations. For instance, tracking algorithms with a better selection of proton tracks were employed to improve the detection efficiency of the $^2$He particle decay. Detection of beam-like particles originating in $(d,{}^{2}\text{He})$ reactions in inverse kinematics provides a better selectivity of the reaction channel and a strong background suppression. In this work, we investigated the use of a kinematic fitting procedure for improving the reconstruction of the reaction parameters. It was shown that the energy resolution of the excited states can be improved by about a factor 2 if beam tracking detectors are used. In the future, the kinematic fitting tool will provide a powerful method for  data analysis in experiments  with complete particle detection. Also, space-charge effects were investigated with a model that accounts for charge build-up and ion recombination. These effects are most relevant around the beam axis, and generate a distortion of the particle tracks in the central region. At relatively high beam intensities and large nuclear charge, the space-charge effects are expected to play a significant role in the deformation of the particle tracks. Therefore, the simulation model presented here was used to generate distortion maps that can be applied in the future to correct experimental data. \par

The analysis tools developed in this work are designed to process both simulated and experimental data. Future $(d,{}^{2}\text{He})$ experiments in inverse kinematics using the AT-TPC will use the code. Many of the routines are also valid for the analysis of other type of reactions involving fast beams such as inelastic scattering and fission.

\section*{Acknowledgments}
This work was supported by the US National Science Foundation under Grants PHY-2209429 (Windows on the Universe: Nuclear Astrophysics at FRIB), PHY-1430152
(JINA Center for the Evolution of the Elements). This material is based upon work supported by the U.S. Department of Energy, Office of Science, Office of Nuclear Physics and used resources of the Facility for Rare Isotope Beams (FRIB), which is a DOE Office of Science User Facility, operated by Michigan State University, under Award Number DE-SC0000661. This work has received financial support from Xunta de Galicia (Centro singular de investigación de Galicia accreditation 2019-2022), by European Union ERDF, and by the “María de Maeztu” Units of Excellence program MDM-2016-0692 and the Spanish Research State Agency. Y. A. acknowledges the support by the Spanish Ministerio de Economía y Competitividad through the Programmes “Ramón y Cajal” with the Grant No. RYC2019-028438-I.

\bibstyle{plain}
\bibliography{references_format}
\end{document}